\documentclass[pre,floatfix,preprint]{revtex4}

\usepackage{epsfig}
\usepackage{times}

\def\be{\begin{equation}}
\def\ee{\end{equation}}

\begin{document}
%\bibliographystyle{revtex}

%\documentclass{ws-ijmpb}

%\begin{document}

%\markboth{Lepp\"anen et al.}
%{Dimensionality effects in Turing pattern formation}

\title{Dimensionality effects in Turing pattern formation}

\author{Teemu Lepp\"anen, Mikko Karttunen, and Kimmo Kaski}
\address{Laboratory of Computational Engineering, 
Helsinki University of Technology,
P.O. Box 9203, FIN--02015 HUT, Finland}

\author{Rafael A. Barrio}
\address{Instituto de Fisica, Universidad Nacional Aut\'onoma de M\'exico, Apartado Postal 20-364, 
01000 M\'exico D.F., M\'exico}

%\begin{history}
%\received{\today}
%\revised{(DAY MONTH YEAR)}
%\end{history}

\begin{abstract}
The problem of morphogenesis and Turing instability are
revisited from the point of view of dimensionality effects. First the 
linear analysis of a generic Turing model is elaborated to the case 
of multiple stationary states, which may lead the system to bistability. 
The difference between two- and three-dimensional pattern formation 
with respect to pattern selection and robustness is discussed. 
Preliminary results concerning the transition between quasi-two-dimensional
and three-dimensional structures are presented and their relation to
experimental results are addressed. 
   
\end{abstract}

%\keywords{Pattern formation, reaction-diffusion system, mathematical biology}
\maketitle
\section{Introduction}

Alan Turing is best known for his contribution to the foundations of 
computer science but he has also played a key role in the birth of 
nonlinear dynamical theory. As he pursued his dream of artificial brain 
he had to concern himself with the problem of biological growth. 
In 1952 he published a paper, where he proposed a mechanism by which 
genes may determine the structure of an organism\cite{turing}. 
His theory did not make any new hypothesis, but it merely suggested 
that the fundamental physical laws can explain many of the emerging 
features. For example in case of a human body the problem is as follows: 
Can $3\times 10^{9}$ base-pairs of DNA code for approximately $10^{11}$ 
neurons, $10^{15}$ synaptic connections and estimated $10^{13}$ cells 
in total? In contrast to a layman's belief that everything, i.e, structure, 
function and behaviour, is determined by genes, it is evident that there 
have to be some quite general physico-chemical mechanisms involved. 

In order to develop a model for this kind of biological growth Turing was 
aware that he had to simplify things a lot, as becomes clear from the 
second sentence of his seminal paper: ``This model will be a simplification 
and an idealization, and consequently a falsification.'' In order to construct 
a manageable model he neglected the mechanical and electrical properties of 
tissue, and instead considered the chemical reactions and diffusion as the 
crucial factors in biological growth for formulating a descriptive theory as
a system of coupled reaction-diffusion equations. Turing showed that this 
kind of system may have a homogeneous stationary state that is unstable 
against perturbations. In fact, any random deviation from this stationary state 
leads to a symmetry break and spatial concentration patterns due to a 
mechanism called diffusion-driven instability. In his paper Turing 
discusses the blastula stage of an embryo, which appears almost spherical 
but shows some deviations from the perfect symmetry to different (random) 
directions in each embryo of a certain species. Hence he stated that they 
can not be of great importance to morphogenesis:  
``From spherical initial state chemical reactions and diffusion certainly 
cannot result in an organism such as a horse, which is not spherically 
symmetrical.'' Later he proved all this to be false because of the random 
deviations in the embryo, and stated: ``It is important that there are 
some deviations, for the system may reach a state of instability 
in which these irregularities tend to grow.''

The fact that all the animals of a certain species do not have exactly 
the same coating patterns supports Turing's idea that there is randomness 
involved in the morphogenesis. For example, all tigers have similar periodic 
patterns of stripes, but the stripes are not in the same exact positions in 
different tigers. Turing assumed the function of genes to be purely catalytic, 
and indeed the genome is known to be the blue-print for the biological 
structure that is to be formed. The genes of a tiger state that it should 
have stripes. As a result, the genes code proteins that set the 
reaction and diffusion rates of the morphogens in such a way that the 
homogeneous state of the chemical system becomes unstable due to local 
or temporal inhomogeneities, which eventually lead to complex pattern 
formation. Here a morphogen stands for a chemical related to biological 
pattern formation.

These reaction-diffusion, or Turing, systems have mostly been 
applied in mathematical biology for explaining pattern formation 
in biological systems\cite{murray}. If one considers the Turing pattern 
of morphogens as a pre-pattern according to which the melanocytes 
(pigment producing cells) in the epidermis differentiate, one could 
explain how some animals get their nonuniform patterns. The problem 
is that the morphogens have not been identified {\it in vivo} and 
thus even their existence can be questioned. Yet, a lot of research 
has been done based on Murray's hypothesis and patterns of 
butterflies\cite{sekimura}, fish\cite{kondo,barrio1}, and lady 
beetles\cite{liaw} have been imitated by numerically solving Turing 
models. Apart from these numerical studies, also dynamic and stability 
aspects of the Turing models have been investigated\cite{judd,lengyel,callahan}.  
In these, one applies the linear stability analysis or more complicated 
nonlinear bifurcation analysis to obtain insight of the system dynamics. 
Typically the parameters of a Turing model are adjusted on the basis of
such an analysis. Despite the simple form of the equations theoretical 
analysis of the time-dependent evolution is very complicated and can only 
be done by analyzing the bifurcation from the stationary state. 
Therefore, the effect of introducing, e.g. inhomogeneous diffusion 
coefficients\cite{barrio2}, growing domains\cite{varea1} or curvature 
of the domain\cite{varea2}, had to be studied numerically. As other
extensions of the Turing systems we have studied the formation of Turing 
structures in three dimensions\cite{teemu}, the effect of noise to 
Turing structures\cite{teemu2} and the dependence of the structural 
characteristics of the morphologies on the parameter selection\cite{teemu3}. 
The first experimental observation of Turing patterns was presented in 1990 
in a single-phase open gel reactor with chloride-iodide-malonic acid (CIMA) 
reaction\cite{castets}. For a review of pattern formation in biochemical
systems, we refer the reader to the extensive article by Hess\cite{hess}.

In the next section of this paper we explain the idea of Turing instability 
in a quite general way. After that we carry out linear analysis in the case 
of multiple stationary states and briefly discuss the possibility of 
bistability in the system. The parameter selection is followed by the 
presentation of numerical results of two- and three-dimensional Turing 
systems. The differences between 2D and 3D systems are considered with 
respect to the pattern selection and robustness against noise. In addition,
preliminary results on the transition between two and three dimensions 
by increasing the thickness of the system are introduced and their 
connection to chemical experiments is discussed. 

\section{Diffusion-driven instability}

Diffusion-driven instability or Turing instability is the mechanism by 
which the random motion of the molecules, i.e., diffusion may make a stable 
state of a chemical system unstable. 
Turing's idea of diffusion-driven instability was ground-breaking for 
the following reasons: 1.) It is counter-intuitive: Typically diffusion 
stabilizes (e.g. a droplet of ink dispersing into water due to 
diffusion). 2.) From any random initial state diffusion-driven instability 
will result in the same morphology according to intrinsic characteristics 
of the system (such as the random deviations in the embryo). 3.) It set 
the basis for research in mathematical biology, nonlinear dynamical theory 
and chemical pattern formation; in 1950s non-equilibrium phenomena, 
symmetry-breaking or complex systems were not fashionable.

The reaction-diffusion mechanism resulting in an instability was
illustrated by Turing with the problem of missionaries and cannibals in 
an island. Missionaries come to the island by boat and want to evangelize 
the cannibals. The rules are as follows: If two or more missionaries 
meet one cannibal they can convert him to a missionary. If the relative 
strength is the other way around, the missionaries get killed and eaten 
by the cannibals. As the missionaries die, more missionaries are brought 
to the island. In addition, cannibals reproduce cannibals. The mechanism for
instability, i.e. diffusion, is in this example the movement of cannibals 
and missionaries. The missionaries are assumed to have bicycles and thus 
they move faster, i.e., they represent the inhibitor of a reaction by 
slowing down the reproduction of cannibals, which in turn are the activator. 
Should the missionaries not have bicycles, they would always get killed 
as they meet cannibals, but by having bicycles they have a chance to escape 
and return when there are more missionaries around. With these definitions 
the auto-catalytic nature of the Turing mechanism becomes evident: In areas 
with a lot of cannibals the number of cannibals will increase due to 
reproduction, then being more effective in killing missionaries. On the other 
hand, the predominance of the cannibals means that more missionaries will be 
brought to the island to convert them. In most cases the cannibals and 
missionaries will finally find a stationary pattern, which corresponds to a 
map of the island where the areas with cannibal dominance can be marked by 
one color and the areas with missionary dominance by another color. 

\section{Formal Turing models}

A Turing model describes the time variation of the concentrations of 
two chemical substances or morphogens due to reaction-diffusion processes. 
Generally a Turing system can be written as
\begin{eqnarray}
U_t & = & D_U \nabla^2 U + f(U,V)\nonumber \\
V_t & = & D_V \nabla^2 V + g(U,V),
\label{eq:turing}
\end{eqnarray}
where $U = U(\vec{r},t)$ and $V = V(\vec{r},t)$ are the spatially varying 
concentrations, and $D_U$ and $D_V$ are the diffusion coefficients of the 
morphogens setting the time scales for the system. The reaction of the 
chemicals is described by the two nonlinear functions $f$ and $g$, which 
in general can be derived from chemical reaction formulas using the law 
of mass action\cite{murray}.

In this paper we use the generic Turing model introduced by Barrio
{\it et al.}\cite{barrio1} for which the reaction kinetics is obtained by
Taylor expanding the nonlinear functions around a stationary solution 
$(U_c,V_c)$ defined by $f(U_c,V_c) = 0$ and $g(U_c,V_c) = 0$.
From Eq.~(\ref{eq:turing}) one can see that this condition, indeed, defines 
the stationary state (time derivative equal to zero) in the absence of 
diffusion. If terms above the third order are neglected, the equations of 
motion read as follows
\begin{eqnarray}
u_t & = & \delta D \nabla^2 u + \alpha u(1-r_1 v^2)  + v(1-r_2 u) \nonumber \\
v_t & = & \delta \nabla^2 v + v(\beta + \alpha r_1 uv) + u(\gamma + r_2 v),
\label{eq:barrio}
\end{eqnarray}
where $u=U-U_{c}$ and $v=V-V_{c}$. Terms $r_1$ and $r_2$ set the amplitudes 
of the nonlinearities, and they describe the reaction, such that quantity 
$r_1$ enhances stripe formation while $r_2$ enhances spots in two 
dimensions\cite{barrio1} and lamellae and spheres in three 
dimensions\cite{teemu}, respectively. $D$ is the ratio of the diffusion 
coefficients of the two chemicals, $\delta$ acts as a scaling factor, 
and $\alpha$, $\beta$ and $\gamma$ contribute to the mode selection.
Here $D \neq 1$ is a necessary, but not sufficient condition for the 
diffusion-driven instability in two or more dimensions\cite{vastano}.

Previously, this model has been studied by setting $\alpha = - \gamma$ 
to keep $(0,0)$ the only stationary solution\cite{barrio1,teemu}. By relaxing 
this condition one can find two more stationary states defined by 
$f(u_c,v_c) = g(u_c,v_c) = 0$, reading as follows
\begin{equation}
u_c^i = \frac{r_2 + (-1)^i \sqrt{r^2_2 + 4 \alpha r_1 (K \beta - \gamma)}}{2 K \alpha r_1}
\label{eq:ss1}
\end{equation}
and
\begin{equation}
v_c^i = - K u^i_c,
\label{eq:ss2}
\end{equation}
where $K = \frac{\alpha+\gamma}{1+\beta}$ and $i = 1,2$.

Sufficient conditions for diffusion-driven instability to occur are 
widely known\cite{murray}. In our case restricting the parameter 
selection such that $\alpha \in (0,1)$, $\beta \in (-1,0)$ and 
$ \gamma \in (-1,0) $ we are left with only two conditions 
\begin{eqnarray}
-1 < \beta < \gamma,\nonumber \\ 
\alpha - 2 \sqrt{D |\gamma|} > \beta D. \nonumber
\end{eqnarray}

Now the dispersion relation of the system (Eq.~(\ref{eq:barrio})) can be 
found by solving the eigenvalues of the linearized system of equations. 
This can be done by substituting a trial solution of the form 
$w(\vec{r},t) = \sum_{k} c_k e^{\lambda t} w_k(\vec{r},t)$ into 
Eq.~(\ref{eq:barrio}), which results in
\[%%\begin{equation}
\left( \begin{array}{c}
\lambda u_k \\
\lambda v_k
\end{array} \right) =
\left( \begin{array}{cc}
D_u & 0 \\
0 & D_v
\end{array} \right)
\left( \begin{array}{c}
-k^2 u_k \\
-k^2 v_k
\end{array} \right) +
\left( \begin{array}{cc}
f_u & f_v \\
g_u & g_v
\end{array} \right)_{u_c,v_c}
\left( \begin{array}{c}
u_k \\
v_k
\end{array} \right),
\]%%\end{equation}
where $f_u$, $f_v$, $g_u$ and $g_v$ denote the partial derivatives of the 
reaction kinetics evaluated at the stationary state $(u_c,v_c)$, which in the  
matrix form are as follows
\begin{eqnarray}
\left( \begin{array}{cc}
f_u & f_v \\
g_u & g_v
\end{array} \right)_{u_c,v_c} =
\left( \begin{array}{cc}
\alpha(1-r_1 v_c^2)-r_2 v_c & 1-r_2 u_c - 2 \alpha r_1 u_c v_c \\
\alpha r_1 v_c^2 + \gamma + r_2 v_c & \beta + 2 \alpha r_1 u_c v_c + r_2 u_c
\end{array} \right),
\label{eq:linear}
\end{eqnarray}
where $u_c$ and $v_c$ are given by Eqs.~(\ref{eq:ss1}) and (\ref{eq:ss2}). 
Now the dispersion relation $\lambda(k)$ can be solved from
\begin{eqnarray}
\lambda^2 + \left[ (D_u+D_v) k^2 - f_u - g_v \right] \lambda + D_u D_v k^4 - k^2 (D_u f_u + D_v g_v) - f_v g_u  =0,
\label{eq:quadr}
\end{eqnarray}
where $k^2 = \vec{k} \cdot \vec{k}$ and for the generic model 
$D_u = \delta D$, $D_v = \delta$. 

If one chooses the parameters such that the two unstable states defined by 
Eqs.~(\ref{eq:ss1}) and (\ref{eq:ss2}) are close to each other, preliminary 
numerical simulations show temporal changes in the concentration patterns
as the system moves between the two stationary states. This bistability 
is most probably due to complex nonlinear coupling of the modes growing 
from the two unstable states. However, quantitative prediction of this 
interesting behavior requires nonlinear bifurcation analysis. The analysis 
and numerical results of the temporal solutions will be presented elsewhere. 
Following Ref.~\cite{teemu}, we fix $\alpha = -\gamma$ from now on.

\begin{figure}[th]
\centerline{\psfig{file=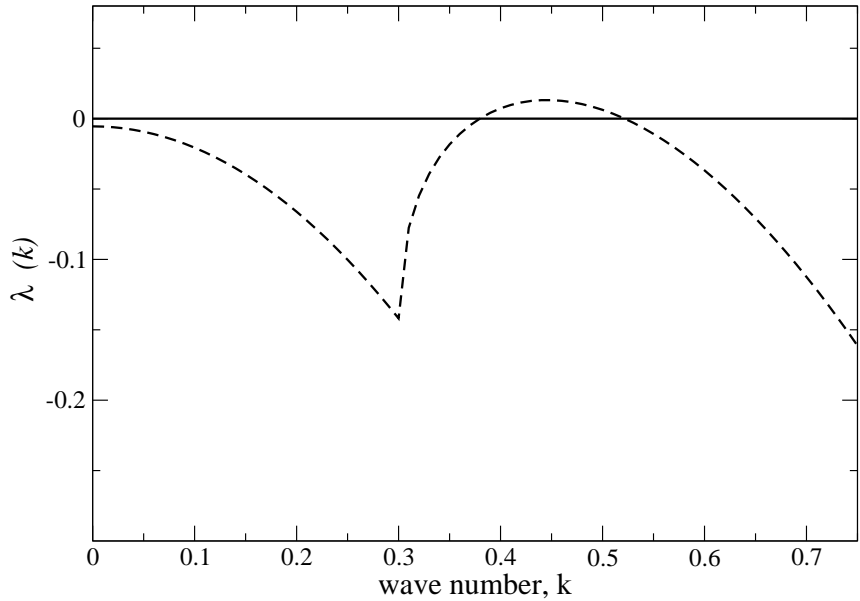, angle=-90, width=.80\textwidth}}
\vspace*{5pt}
\caption{The real part of dispersion relation for the critical wave 
vector $k_c=0.45$ given by Eq.~(\ref{eq:dispersion}). The unstable modes are 
all $k$~for which $\lambda (k)> 0$.}
\label{fig:dispersion}
\end{figure}

By observing the boundaries of the region with positive growth rate, i.e., 
$\lambda (k)> 0$ one can analytically derive the modulus of the critical 
wave vector
\[%% \begin{equation}
k_c^2 = \frac{1}{\delta} \sqrt{\frac{\alpha (\beta +1)}{D}}.
\]%% \end{equation}
By adjusting the parameters and allowing only a few modes to be unstable 
one can end up with several different parameter sets. Here we use the 
parameters $D=0.516$, $\alpha =0.899$, $\beta = -0.91$ and $\delta = 2$,
which correspond to critical wave vector $k_c=0.45$ and which we have 
used earlier\cite{teemu}. These selections fix the characteristic length of 
the pattern to be $2 \pi / k_c$. The corresponding dispersion relation can 
be calculated based on Eq.~(\ref{eq:quadr}), which in the case of 
$u_c = v_c = 0$ reduces to
\begin{equation}
\lambda^2 - (\alpha +\beta - \delta k^2 (1+D)) \lambda + (\alpha -\delta D k^2)(\beta - \delta k^2) 
- \gamma=0.
\label{eq:dispersion}
\end{equation}
The real part of this dispersion relation is plotted in Fig.~\ref{fig:dispersion}.
For a more detailed discussion, see Ref.~\cite{teemu}.

\section{Numerical results}

The numerical simulations were carried out by discretizing the system 
into a spatial mesh or lattice. The Laplacian was calculated by finite 
difference method, with $dx = 1.0$, and the equations were iterated in 
time using Euler's scheme with time step $dt = 0.05$. In a discretized 
three-dimensional system, the wave number values are not continuous 
but of the form
\begin{equation}
|\vec{k}|= 2 \pi \sqrt{\bigg(\frac{n_x}{L_x}\bigg)^2 + \bigg(\frac{n_y}{L_y}\bigg)^2 + \bigg(\frac{n_z}{L_z}\bigg)^2},
\label{eq:wavevector}
\end{equation}
where $L_{x,y,z}$ are the system dimensions to the corresponding directions 
and $n_{x,y,z}$ are the wave number indices. In this paper we have used periodic 
boundary conditions for the concentration fields, which were initially set as
random perturbations around the stationary state, i.e., random numbers 
with zero mean and variance of $\sim 0.05$.

\subsection*{2D patterns and 3D structures}

\begin{figure}[th]
\centerline{\psfig{file=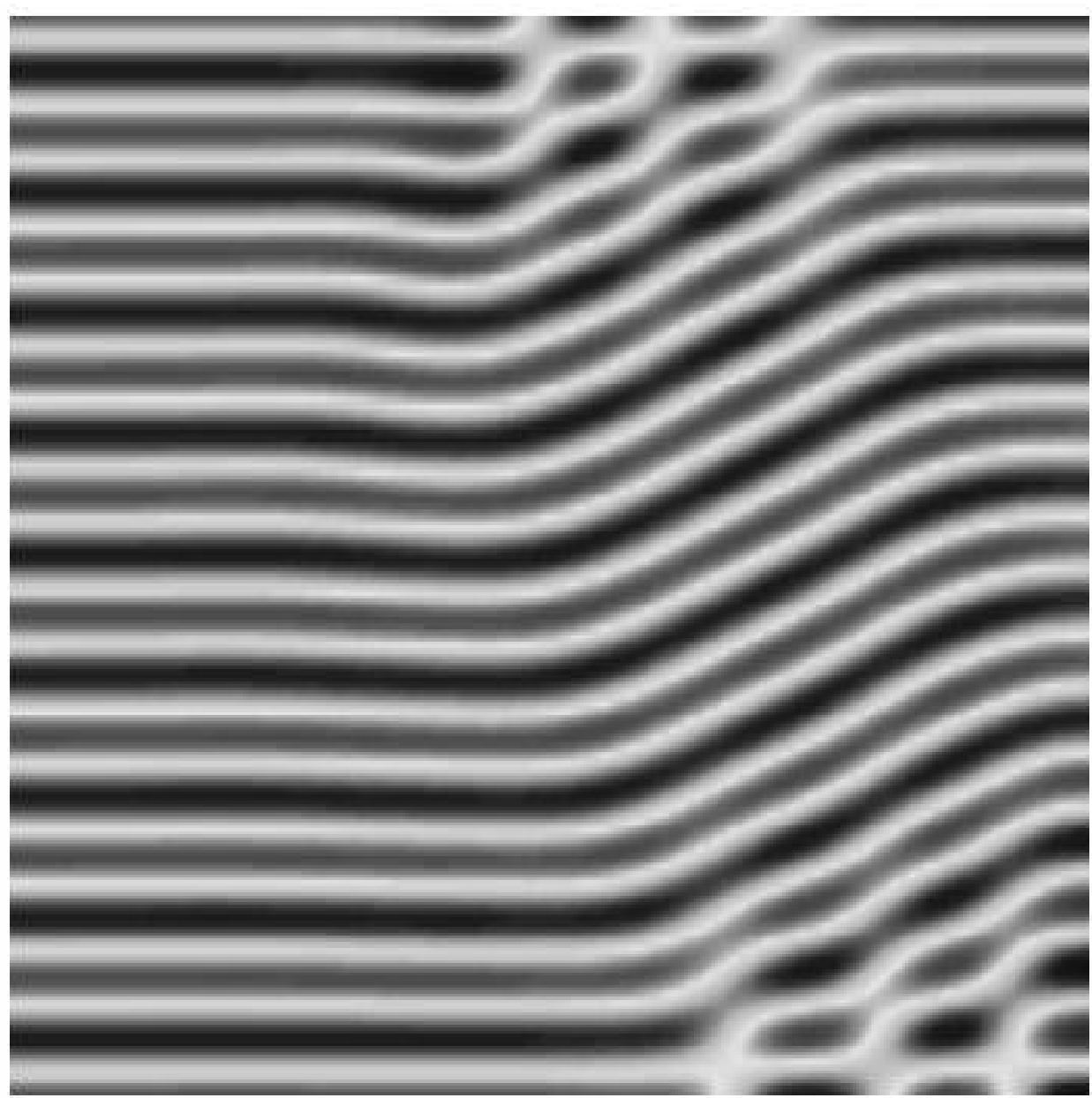, angle=90, width=.4\textwidth}
\psfig{file=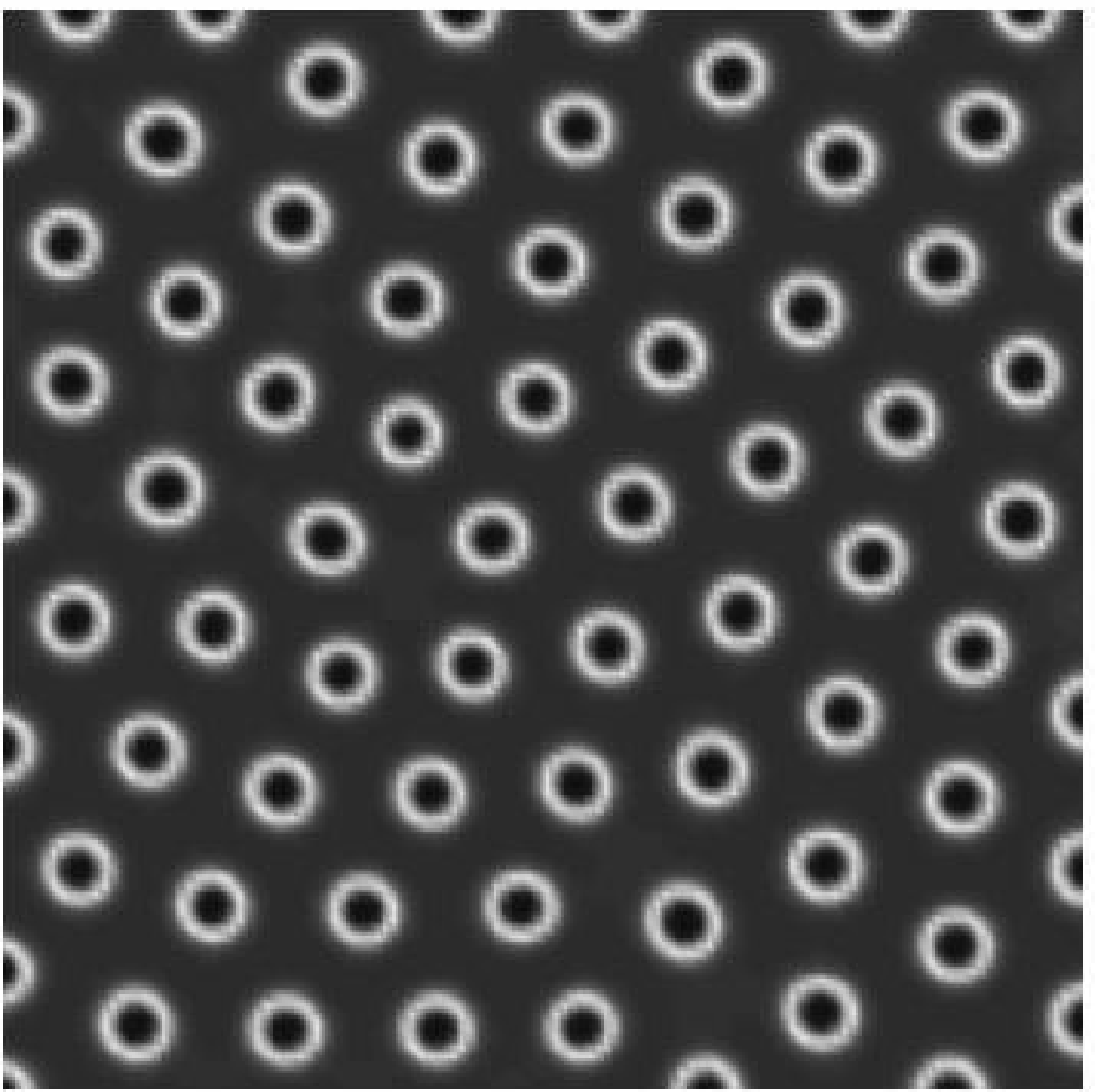, angle=0, width=.4\textwidth}}
\vspace*{5pt}
\caption{The 2D patterns in a system of size $128 \times 128$ obtained 
from the numerical simulation of Eq.~(\ref{eq:barrio}). The parameters 
correspond to the mode $k_c = 0.45$. Left: $r_1 = 3.5$ and $r_2 = 0$. 
Right: $r_1 = 0.02$ and $r_2 = 0.2$.}
\label{fig:2d}
\end{figure}

In the simulations the characteristic length of the pattern is fixed by 
the parameter selection and the morphology of the resulting pattern can 
be adjusted by the nonlinear parameters $r_1$ and $r_2$. The coefficient 
$r_1$ of the cubic term enhances stripe or lamellae formation, whereas 
$r_2$ enhances the formation of spherical shapes. Figure~\ref{fig:2d} 
shows the results of numerical simulations for two-dimensional 
$128 \times 128$ systems, where the nonlinearities favour either spots 
or stripes.

\begin{figure}[th]
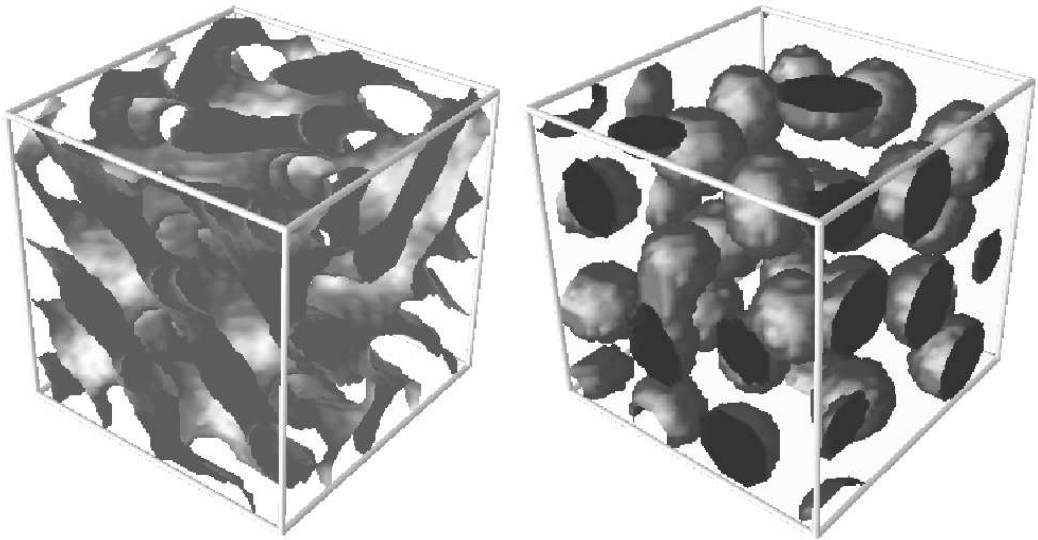

%\centerline{
\psfig{file=3D_stripes2.epsf, angle=0, width=.425\textwidth}%%\\
\psfig{file=3D_spots2.epsf, angle=0, width=.425\textwidth}
%}
\vspace*{5pt}
\caption{The 3D structures in a system of size $50 \times 50 \times 50$
obtained from the numerical simulation of Eq.(~\ref{eq:barrio}).
The structure is visualized by plotting the iso-surface for one of the 
concentration fields. The parameters correspond to the mode $k_c = 0.45$. 
Left: $r_1 = 3.5$ and $r_2 = 0$. Right: $r_1 = 0.02$ and $r_2 = 0.2$.}
\label{fig:3d}
\end{figure}

Extending the pattern formation problem from 2D to 3D is by no 
means straightforward. Figure~\ref{fig:3d} shows 3D structures 
corresponding to the patterns of Fig.~\ref{fig:2d}. The two-dimensional 
stripes become complex and aligned lamellae in the three-dimensional 
system instead of pure lamellar planes that one would expect 
(see Fig.~\ref{fig:noise_planes}). This is due to one more degree 
of freedom. Planes are formed, but the resulting structure is, as a 
matter of fact, a combination of aligned planes crossing each 
other. The system dynamics is unable to organize the three-dimensional 
structure into a more regular shape. The complex lamellar structure 
satisfies the symmetry requirements imposed by the nonlinearities 
though the resulting structure is not optimal. In the case of 
enhanced quadratic nonlinear interaction one obtains 3D spherical 
shapes as one would expect (Fig.~\ref{fig:3d}), but the packing 
is not FCC. 

\subsection*{Dimensionality transition}

The transition from 2D patterns to 3D structures is a more challenging 
problem than the direct comparison of the final results (Figs.~\ref{fig:2d} 
and \ref{fig:3d}). It has been shown experimentally that an open gel 
reactor with CIMA reaction may have a bistability, i.e., both spot and 
stripe patterns may appear into the gel on different heights. This has 
been explained by a concentration gradient that is imposed by the 
reactor\cite{quyang}. Turing patterns have also been studied in ramped 
systems, where the thickness of the gel is increased gradually. In that 
case one observes qualitatively different patterns corresponding at 
different thicknesses\cite{dulos}. 

The problem has also been addressed more quantitatively by Dufiet 
and Boissonade\cite{dufiet}. They modeled the pattern formation of a 
three-dimensional experimental reactor by imposing a permanent 
gradient on one of the bifurcation parameters of their model. This 
corresponds to the fact that in an experimental reactor the concentrations 
are kept constant only on the feed surfaces, whereas the concentration 
inside the gel is governed by reaction and diffusion. They claim that 
the patterns in quasi-2D reactors can, to a certain extent, be interpreted 
as two-dimensional patterns when the thickness of the gel~$L_z$ is less 
than the characteristic wavelength $\lambda_c = 2 \pi / k_c$~of the pattern.

\begin{figure}[th]
\centerline{
\psfig{file=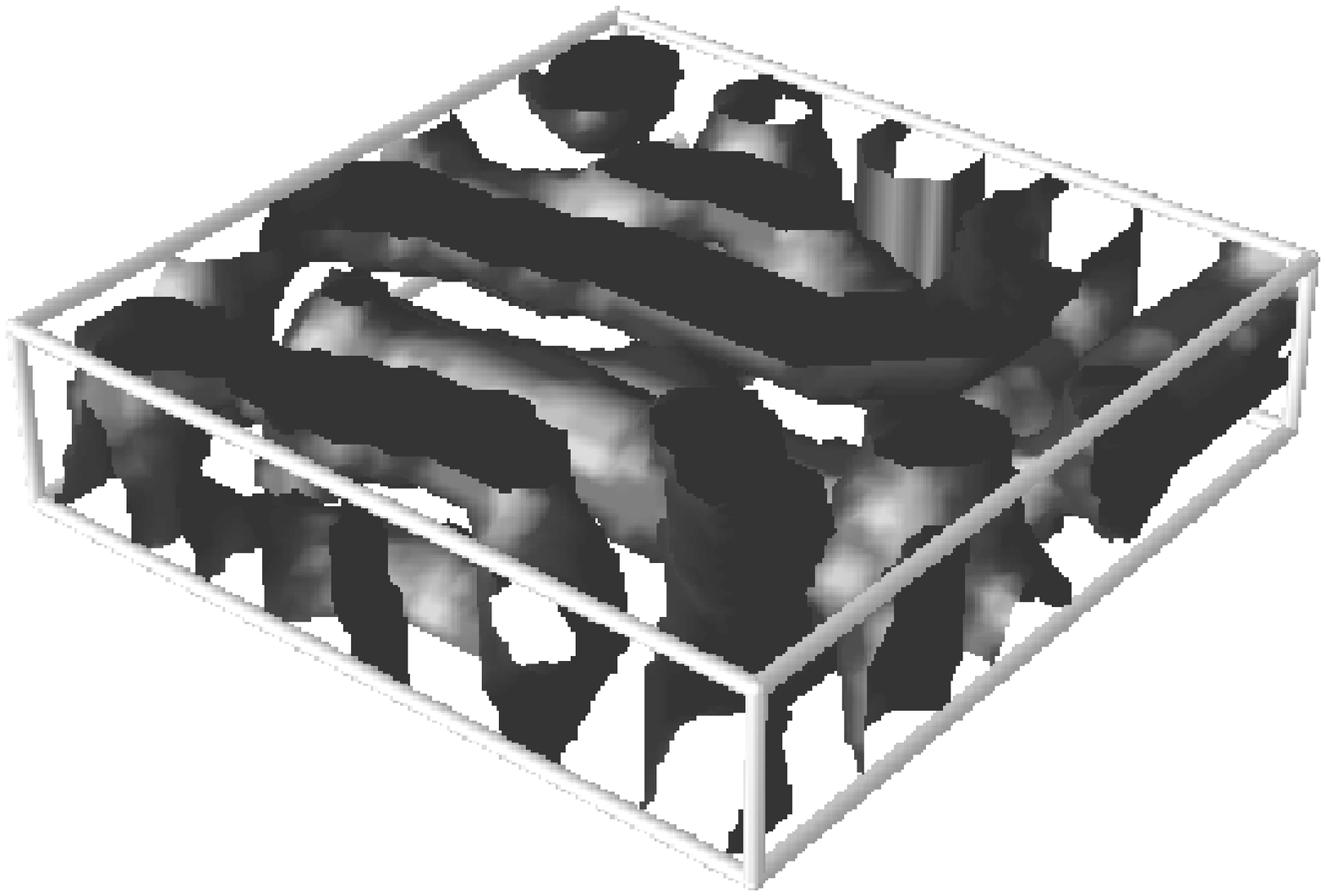, angle=0, width=.425\textwidth}
\psfig{file=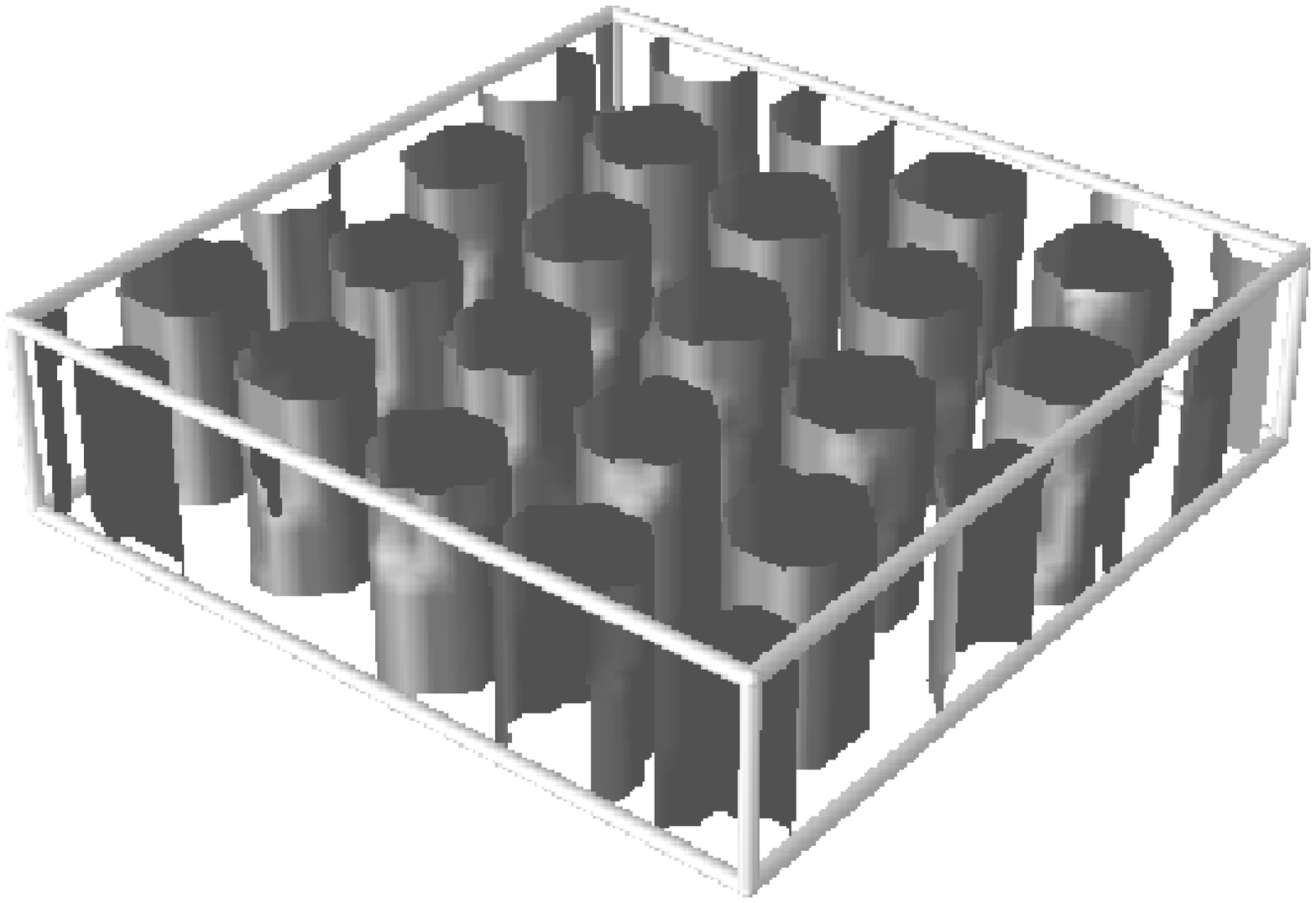, angle=0, width=.425\textwidth}}
\vspace*{5pt}
\caption{Structures obtained from numerical simulation of the 
Lengyel-Epstein model (Eq.~(\ref{eq:le}))
in a domain of size $100x100xL_z$. The dimensionality of the pattern is not 
determined by the thickness $L_z$ of the system alone. Left: $L_z=12$. 
Right: $L_z=13$. The system parameters are the same for both 
%%($\lambda_c = 6.86$), 
but  the random initial conditions are different.}
\label{fig:transition}
\end{figure}

As for the effect of dimensionality in the generic Turing model of 
Eq.~(\ref{eq:barrio}) we have observed that the transition from a 2D to a 3D 
system is not as simple as it has been thought. There seems to be some 
stochastic behaviour in the transition: As one increases the thickness of 
the system, it will lose the correlation between the bottom and top plates 
gradually, but the transition thickness is not well-defined, i.e., 
$L_z > \lambda_c$ does not guarantee that the structure becomes 
three-dimensional. Figure~\ref{fig:transition} shows the resulting structures 
for two systems of nearly the same thickness $L_z > \lambda_c$ as one starts 
from random initial conditions. One can easily observe that the leftmost 
structures is three-dimensional, whereas the rightmost is quasi-two-dimensional, 
although the latter structure grows in a thicker domain. The preliminary 
results presented in Fig.~\ref{fig:transition} were obtained by straightforward 
simulations of the Lengyel-Epstein model\cite{lengyel}, as follows 
\begin{eqnarray} 
u_t & = & \frac{1}{\sigma}(a-u-4\frac{uv}{1+u^2} + \nabla^2 u) \nonumber \\
v_t & = & b (u-\frac{uv}{1+u^2})+ d \nabla^2 v,
\label{eq:le}
\end{eqnarray}
where $a$, $b$, $d$ and $\sigma$ are adjustable parameters. 
We used the parameter set $\sigma = 50$, $d = 1.07$, 
$a=8.8$ and $b=0.09$, which is known to correspond
to hexagonal patterns\cite{rudovics}.

Based on our numerical studies using both the generic Turing model
(Eq.~(\ref{eq:barrio})) and the Lengyel-Epstein model (Eq.~(\ref{eq:le}))
it seems that the probability that the pattern is three-dimensional 
is proportional to the number of wave vectors $\vec{k}$ (Eq.~(\ref{eq:wavevector})) 
with $n_z \neq 0$ such that the mode $k$ is unstable according to the 
dispersion relation (Eq.~(\ref{eq:dispersion})). Our results show that 
the number of three-dimensional systems does not change continuously 
as a function of the system thickness, but the dependence is very 
complicated. It should also be noted that linear analysis is not 
sufficient to explain the complex dynamics of the dimensionality 
transition. Thus some sort of non-linear analysis seems necessary.
However, from the numerical simulations one can calculate interesting 
statistics of the morphologies (e.g. correlations, structure factors), 
although this kind of structural analysis is difficult to carry out for 
chemical patterns in a gel. In addition visualization of numerical results 
makes it possible to see into the structure. We have used the Lengyel-Epstein 
model reaction\cite{lengyel,rudovics} to analyze further the structures 
that have been reported from experiments in ramped systems\cite{dulos}. 
It seems that the experimental patterns can be easily misinterpreted, 
because the depth information is lost in the 2D projection made for 
visualization. What seems to be a non-harmonic modulation may be just 
an aligned lamellae, and what seems to be a combination of stripes and 
spots may as well be tubes, which appear to the observer as stripes 
if seen from the side and as spots if seen from the end. 

As discussed above an important feature of experiments, which is absent 
in the numerical simulations is the gradients in the reactor. 
In a computer a given model can be solved to the computational precision, 
from which it follows that if a model captures the phenomenon accurately, 
the results are accurate and there are no such artifacts as concentration 
gradients. If one thinks of applying Turing system in biological modeling, 
it is of great importance that the models imitate the real processes and 
there is nothing that skews the results. In biological tissue
the sources controlling the parameters of the reaction would in 
part be in the cells and not always in the boundaries. Thus the 
gradients are not always present in biological morphogenesis, from 
which it follows that the patterns obtained from numerical simulations 
may in fact be more realistic than the patterns arising in experimental 
reactors.

\subsection*{The effect of noise on lamellar structures}

We have recently reported results of the general Turing system 
concerning the effect of noise on spherical structures\cite{teemu2}, 
which are more robust against noise than stripes or lamellar structures. 
Here we present results on the effect of noise specifically in the 
presence of cubic nonlinear interaction, which favours stripes or lamellae. 
In order to study this effect we introduce uncorrelated Gaussian noise 
sources $\eta(\vec{x},t)$ such that the equations of motion read as follows
\begin{eqnarray}
u_t & = & D \delta \nabla^2 u + f(u,v) + \eta_u \nonumber \\
v_t & = & \delta \nabla^2 v + g(u,v) + \eta_v,
\label{eq:turing-noise}
\end{eqnarray}
where the first and the second moment of the noise is defined as 
$\langle \eta(\vec{x},t) \rangle = 0$ and 
$\langle \eta(\vec{x},t) \eta(\vec{x}',t') \rangle = A^2 \delta(\vec{x}-\vec{x}') \delta(t-t')$, with the angular bracket denoting the average and $A$ the noise 
intensity. It is noted that the noise was added to each lattice site
at every time step of the simulation. Due to discretization the noise 
has to be normalized such that 
\begin{equation}
\eta = \frac{A}{(dx)^{d/2} \sqrt{dt}},
\end{equation}
where $d$ is the dimension of the system, $dx$ the lattice constant 
and $dt$ the time step.

\begin{figure}[th]
\centerline{\psfig{file=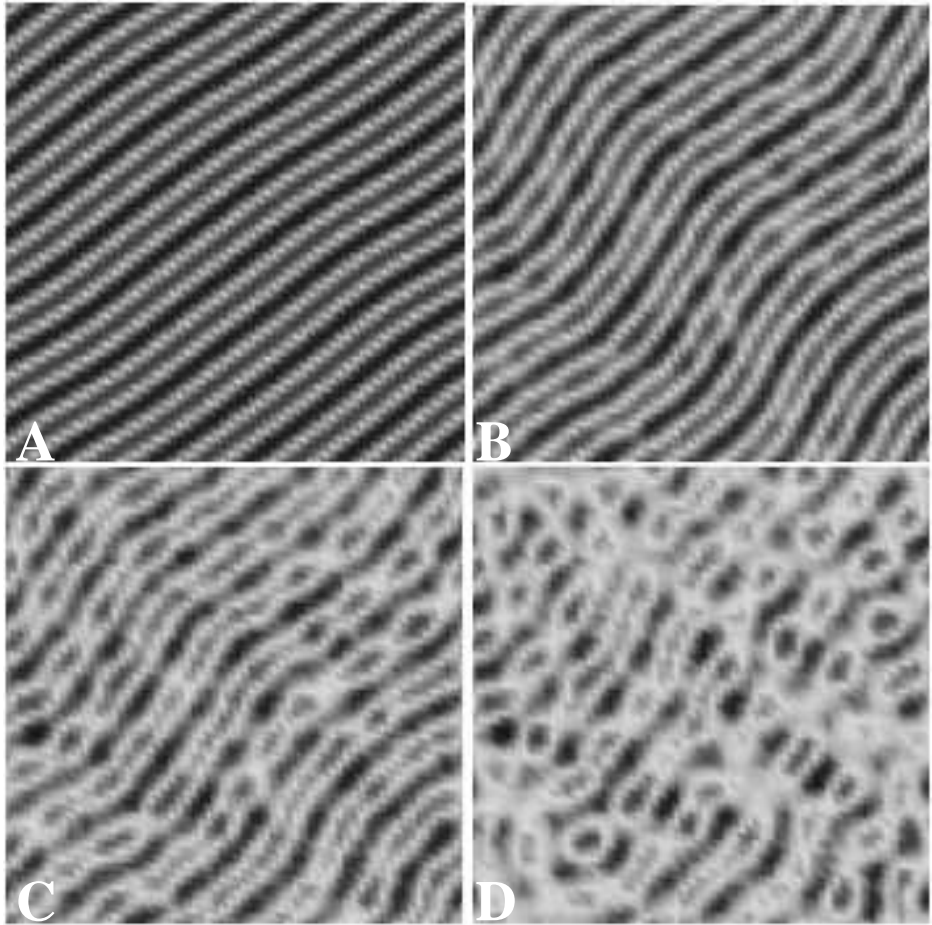, angle=0, width=.75\textwidth}}
\vspace*{5pt}
\caption{Two-dimensional striped pattern in a system of size $64 \times 64$.
The amplitude of the noise is A) $A = 0$, B) $A = 0.02$, C) $A = 0.04$ 
and D) $A = 0.06$ corresponding approximately to $0-40\%$ of the 
amplitude of the modulated concentration wave.}
\label{fig:noise_stripes2d}
\end{figure}

In Figure~\ref{fig:noise_stripes2d} we show the two-dimensional patterns 
corresponding to different noise intensities for nonlinear parameters 
$r_1 = 3.5$ and $r_2 = 0$. The noise was applied all the way throughout 
the evolution, and one can observe that the stripes fade into the noise 
gradually. It should be noted that there are complex dynamics involved 
in the evolution, but the patterns are still formed according to the 
system parameters, and even a substantial noise does not distort the 
evolution completely, but only makes the resulting patterns look noisy. 

\begin{figure}[th]
\centerline{\psfig{file=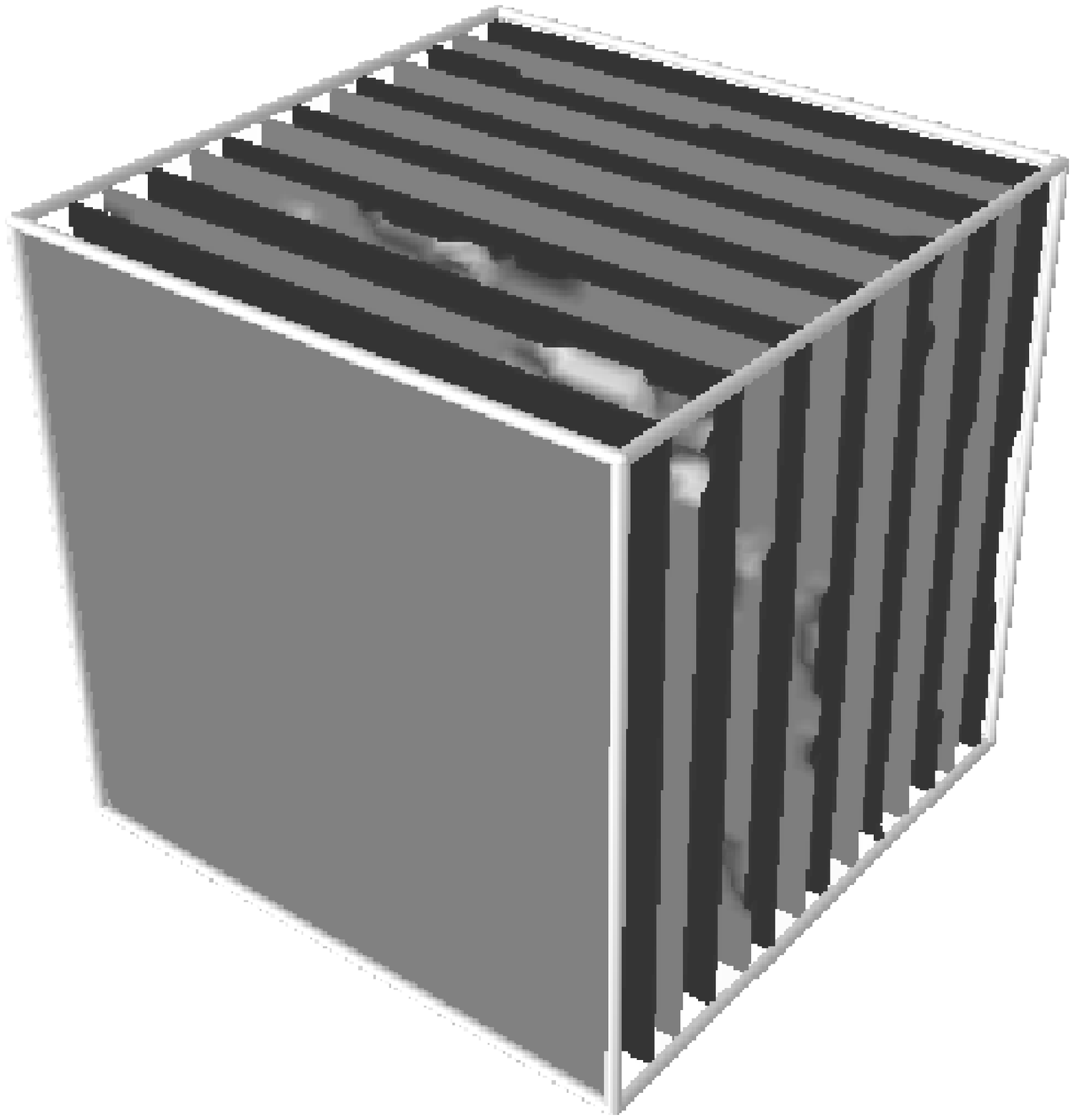, angle=0, width=.425\textwidth}
\psfig{file=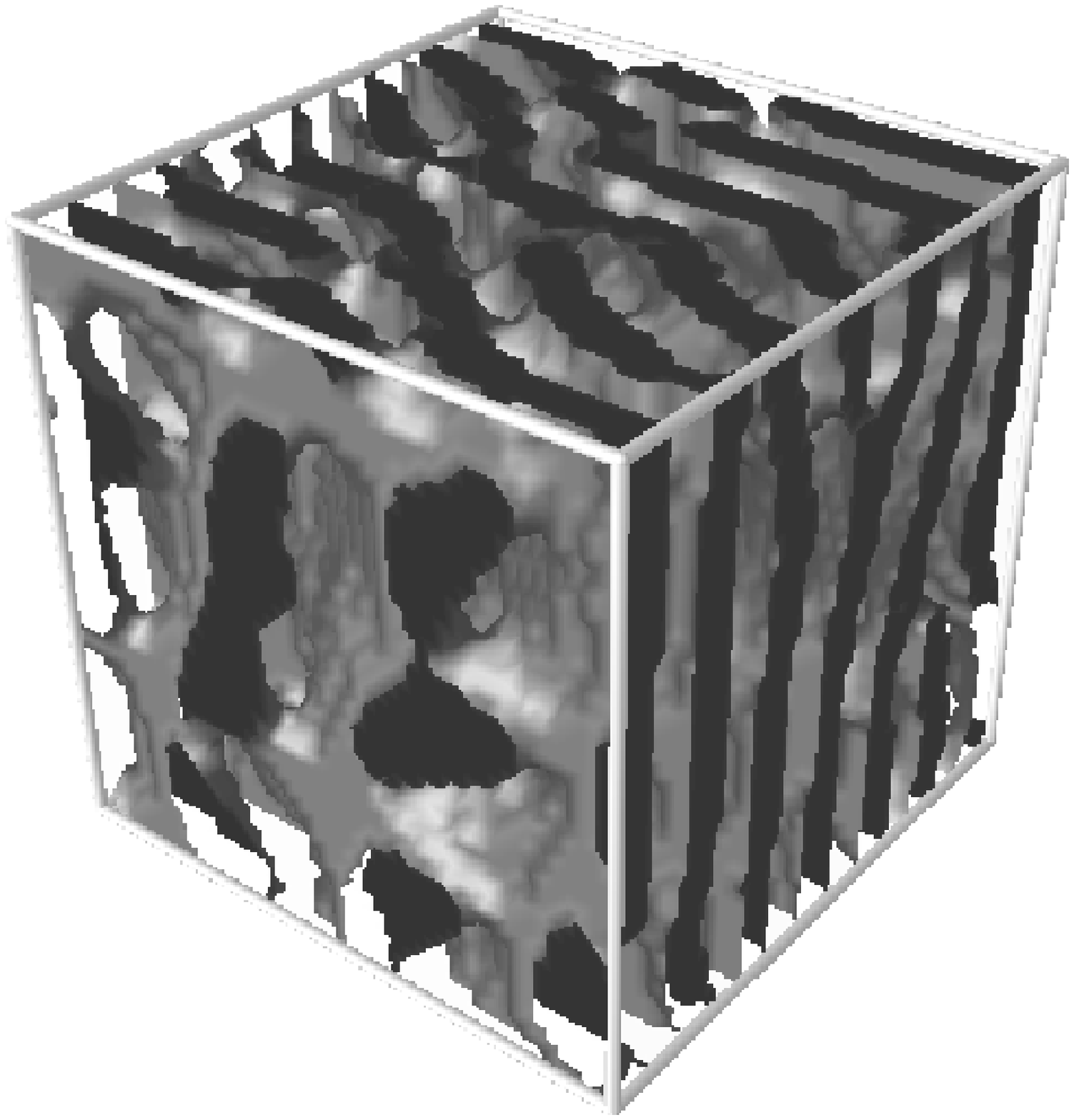, angle=0, width=.425\textwidth}}
\vspace*{0.5cm}
\caption{Three-dimensional planar structure in a system of size 
$50 \times 50 \times 50$. The structure on the left is without noise 
and the one on the right with the noise amplitude $A = 0.02$,
which corresponds approximately to $10\%$ of the amplitude of the 
modulated concentration wave.}
\label{fig:noise_planes}
\end{figure}

We have shown earlier that by using a special initial condition in the 
mid-plane of the system, one can obtain a three-dimensional structure of 
pure parallel planes\cite{teemu}. Figure~\ref{fig:noise_planes} shows such a 
planar structure and the same structure in the presence of noise. One can 
clearly see that the planes get holes on them and align even in the 
presence of a small amount of noise. The noise intensity in the rightmost 
structure of Fig.~\ref{fig:noise_planes} corresponds to that of 
Fig.~\ref{fig:noise_stripes2d}B, where the stripes are intact. This 
explains the fact that the three-dimensional pattern selection is 
very sensitive and the pure planes are not obtained from random 
initial conditions.

In general, we find that the three-dimensional case is much more robust against 
noise than the two-dimensional one, as can be observed with the help of a 
structure factor analysis\cite{teemu2}. The interfaces of the chemical domains 
become distorted, but one can still localize the structure even for noise 
intensities up to 50\% of the amplitude of the chemical concentration 
wave. Such high level robustness of Turing patterns is very interesting as such 
but also from the point of view of applying Turing systems to biological growth,
in which morphogenesis seem to sustain well against external random distortions.

\section{Discussion}

In this paper, we have studied the characteristics of two-dimensional, 
quasi-two-dimensional and three-dimensional pattern formation in Turing 
systems. Our preliminary results indicate that the system may behave as a 
three-dimensional system for certain thickness and as one {\it increases} 
the thickness, the system may behave quasi-two-dimensionally with the same 
set of parameters. Also we find that the pattern selection in a 3D 
system is much more complex than in 2D. Although spherical structures 
are analogous to spotty patterns, similar generalization from stripes to 
lamellae is not as easy. A simulation of a three-dimensional system in 
the presence of a cubic nonlinear interaction results in aligned lamellae with 
holes. This may be due to the fact that it is more difficult to maintain 
long-range order in three dimensions, which is required for lamellae without 
holes. One can observe the breaking of long-range order also in the case of 
spherical structures: The two-dimensional spots are organized in an almost 
perfect triangular lattice, whereas in 3D spherical structures seem to be 
unable to establish regular ordering.

In the study of the effect of noise on striped and planar structures we found 
that these structures and the dynamics of the instability are proportionally 
speaking quite robust against random perturbations. However, spherical 
structures can sustain more noise in absolute values since the amplitude 
of the concentration wave of a spherical structure is larger. We have also 
shown that the three-dimensional planar structure did not sustain noise as 
well as the corresponding two-dimensional stripe pattern although 
three-dimensional structures should be more robust than two-dimensional 
patterns as indicated by our preliminary studies.

In this paper we have also found that the dimensionality can affect the pattern 
formation process in a very profound way and thus understanding the differences 
between two-dimensional pattern formation and three-dimensional structure 
formation is of great importance. The real biological processes are always 
taking place in a three-dimensional domain, although the symmetry restrictions, 
instability and dynamics governing the process may actually be 
quasi-two-dimensional. Therefore, choosing whether a system should be treated 
as three-dimensional or quasi-two-dimensional is by no means trivial and 
thus needing further theoretical and numerical studies for deeper understanding 
of morphogenesis modeling.

\section*{Acknowledgements}

One of us (R.\,A.\,B.) wishes to thank the Laboratory of Computational 
Engineering at Helsinki University of Technology for their hospitality. 
This work has been supported by the Finnish Academy of Science and 
Letters (T.\,L.). and the Academy of Finland through its Centre of 
Excellence Program (T.\,L. and K.\,K.). 

\vspace*{0.5cm}

Animations of Turing systems are available at\\
{\tt http://www.lce.hut.fi/research/polymer/turing.shtml}


\begin{thebibliography}{0}

%%%%%%%%%%%%%%%%%%%%%%%%%%%%%%%%%%%%%%%%%%%%%%%%%%%%%%%%%%%%
% Command and Example                                      %
%                                                          %
% \bibitem{REFERENCE_LABEL} AUTHORS NAMES,                 %
% {\it JOURNAL"S NAMES}{\bf VOLUME NUMBER}, PAGE (YEAR).   %
%                                                          %
% Three examples given below.                              %
%%%%%%%%%%%%%%%%%%%%%%%%%%%%%%%%%%%%%%%%%%%%%%%%%%%%%%%%%%%%

\bibitem{turing} A.M. Turing, {\it Phil. Trans. R. Soc. Lond.}
{\bf B237}, 37-72 (1952).

\bibitem{murray} J.D. Murray, {\it Mathematical Biology}, 2nd. ed., 
(Springer Verlag, Berlin 1993). 

\bibitem{sekimura} T. Sekimura, A. Madzvamuse, A.J. Wathen, and
P.K. Maini, {\it Proceedings Royal Society London B} {\bf 267}, 851 (2000).
   
\bibitem{kondo} S. Kondo, and R. Asai, {\it Nature} {\bf 376}, 678  (1995). 
  
\bibitem{barrio1} R.A. Barrio, C. Varea, J.L. Arag\'on, and P.K. Maini, 
{\it Bull. Math. Biol.} {\bf 61}, 483 (1999). 

\bibitem{liaw} S.S. Liaw, C.C. Yang, R.T. Liu, and J.T. Hong,
{\it Phys. Rev. E} {\bf 64}, 041909 (2002).

\bibitem{polish} A.L. Kawczynski and B. Legawiec, {\it Phys. Rev. E} {\bf 64}, 
056202 (2001).
 
\bibitem{judd} S.L. Judd and M. Silber, {\it Physica D} {\bf 136}, 45 (2000).

\bibitem{lengyel} I. Lengyel and I. R. Epstein, {\it Proc. Nat. Acad. Sci.} 
{\bf 89}, 3977 (1992).

\bibitem{callahan} T.K. Callahan and E. Knobloch, {Physica D} {\bf 132}, 
339 (1999).

\bibitem{barrio2} R.A. Barrio, J.L. Arag\'on, M. Torres, and P.K. Maini, 
{\it Physica D} {\bf 168-169}, 61 (2002).

\bibitem{varea1} C. Varea, J.L. Arag\'on, and R.A. Barrio, {\it Phys. Rev. E}
{\bf 60}, 4588 (1999).

\bibitem{varea2} C. Varea, J.L. Arag\'on, and R.A. Barrio, {\it Phys. Rev. E}
{\bf 56}, 1250 (1997).

\bibitem{teemu} T. Lepp\"anen, M. Karttunen, K. Kaski, R.A. Barrio, 
and L. Zhang, {\it Physica D} {\bf 168-169}, 35 (2002).

\bibitem{teemu2}  T. Lepp\"anen, M. Karttunen, R.A. Barrio, and K. Kaski,
The effect of noise on Turing patterns, submitted 2002.

\bibitem{teemu3}  T. Lepp\"anen, M. Karttunen, R.A. Barrio, and K. Kaski, 
Connectivity of Turing structures, submitted 2003. cond-mat/0302101.

\bibitem{castets} V. Castets, E. Dulos, J. Boissonade and P. de Kepper, 
{\it Phys. Rev. Lett.} {\bf 64}, 2953 (1990). 

\bibitem{hess} B. Hess, 
%%{\it Periodic patterns in biochemical systems},
{\it Quarterly Rev. Biophys.} {\bf 30}, 121 (1997).

\bibitem{vastano} J.A.~Vastano, J.E.~Pearson, W.~Horsthemke, and H.L.~Swinney, 
{\it Phys.~Lett.~A} {\bf 124}, 6 (1987).

\bibitem{quyang} Q. Quyang, Z. Noszticzius, and H.L. Swinney,
{\it J. Phys. Chem.} {\bf 96}, 6773 (1992).

\bibitem{dulos} E. Dulos, R. Davies, B. Rudovics, and P. de Kepper,
{\it Physica D} {\bf 98}, 53 (1996).

\bibitem{dufiet} V. Dufiet and J. Boissonade, {\it Phys. Rev. E} {\bf 53}, 
4883 (1996).

\bibitem{rudovics} B. Rudovics, E. Barillot, P.W. Davies, E. Dulos, 
J. Boissonade, and P. de Kepper, {\it J. Phys. Chem.} {\bf 103}, 1790 (1999).

\end{thebibliography}
\end{document}